\documentclass[11pt]{article}
\usepackage{amsmath}
\usepackage{amsfonts}
\usepackage{amssymb}
\usepackage{graphicx}
\usepackage{bm}
\usepackage{color}
\usepackage{amscd}

\def\bea{\begin{eqnarray}}
\def\eea{\end{eqnarray}}

\begin{document}
\begin{center}
\LARGE {\bf Quasi-Local Conserved Charges of Spin-3 Topologically Massive Gravity }
\end{center}

\begin{center}
{M. R. Setare \footnote{E-mail: rezakord@ipm.ir}\hspace{1mm} ,
H. Adami \footnote{E-mail: hamed.adami@yahoo.com}\hspace{1.5mm} \\
{\small {\em  Department of Science, University of Kurdistan, Sanandaj, Iran.}}}\\

\end{center}

\begin{center}
{\bf{Abstract}}\\
In this paper we obtain  conserved charges of spin-3 topologically massive gravity by using a quasi-local formalism.
We find a general formula to calculate conserved charge of the spin-3 topologically massive gravity which
 corresponds to a Killing vector field $\xi$.
We show that this general formula reduces to the previous one for the ordinary spin-3 gravity presented in \cite{31} when we take into account only transformation under diffeomorphism, without considering generalized Lorentz gauge transformation (i.e. $\lambda _{\xi} =0$), and by taking $\frac{1}{\mu}\rightarrow 0$. Then we obtain a general formula for the entropy of black hole solutions of the spin-3 topologically massive gravity. Finally we apply our formalism to calculate energy, angular momentum and entropy of a special black hole solution and we find that obtained results are consistent with previous results in the limiting cases. Moreover our result for energy, angular momentum and entropy are consistent with the first law of black hole mechanics.

\end{center}

\section{Introduction}

Higher spin gravity was formulated by Vasiliev  and collaborators in papers \cite{0} . In its simplest
form it is an extension of ordinary gravity that includes a
massless scalar and massless fields with spins $S=3,4,...$. In \cite{00} Vasiliev proposed a system of gauge invariant nonlinear dynamical equations for totally symmetric massless fields of all spins in (A)dS backgrounds.
 According to the result of this paper, "in the framework of gravity, unbroken higher spin gauge symmetries require a non-zero cosmological constant".
In paper \cite{3} the authors have considered the coupling of a symmetric spin-3 gauge field $\varphi _{\mu \nu \lambda}$ to 3-dimensional
gravity in a second order metric-like formulation.
 In the context of frame-like approach the gravitational coupling of a symmetric tensor of rank 3 in the presence of negative cosmological constant  can be given by $SL(3, \mathbb{R})\times SL(3, \mathbb{R})$ Chern-Simons theory \cite{1,2',2''}. The emergence of $W$-algebras as asymptotic symmetries of higher-spin gauge
theories coupled to three-dimensional Einstein gravity with a negative cosmological constant
 has been discussed in \cite{1}.\\
Higher-spin theories in $AdS_3$, like ordinary gravity, they possess no propagating degrees of freedom \cite{1,3'}.
 Pure Einstein-–Hilbert gravity in three dimensions exhibits no propagating physical degrees of freedom \cite{4',5'}.
But adding the gravitational Chern-Simons term produces a
propagating massive graviton \cite{6'}. The resulting theory
is called topologically massive gravity (TMG).
The authors of \cite{12} have done the generalization of topologically massive gravity to
higher spins, specifically spin-3 (see also \cite{7'}).
In this paper we would like to obtain conserved charges of black hole solutions of spin-3 topologically massive
 gravity by using a quasi-local formalism. Here, we just consider diffeomorphism covariance and we find energy,
 angular momentum, and entropy,  but we don't investigate the higher-spin charges.\\
A method to calculate the energy of asymptotically AdS solution was given by Abbott and Deser \cite{9'}. Deser and Tekin have extended this approach to the calculation of the energy of asymptotically dS or AdS solutions in higher curvature gravity models and also to TMG \cite{10'}.  The authors of \cite{16} have obtained the quasi-local conserved charges for  black holes in any diffeomorphically invariant theory of gravity. By considering an appropriate variation of the metric, they have established a one-to-one correspondence between the ADT approach and the linear Noether expressions. They have extended this work to a theory of gravity containing a gravitational Chern-Simons term in \cite{17}, and have computed the off-shell potential and quasi-local conserved charges of some black holes in TMG. In the first order formalism, quasi-local conserved charges of Lorentz-diffeomorphism covariant theories of gravity are investigated in \cite{29}. But there are theories which are not Lorentz-diffeomorphism covariant so for such theories a
method for which one can calculate conserved charges of
Lorentz-diffeomorphism covariant theories are useless. In previous paper \cite{9} by introducing the total variation of a quantity due to an infinitesimal Lorentz-diffeomorphism transformation, we have obtained the conserved charges in Lorentz-diffeomorphism non-covariant
theories. The formalism proposed in \cite{9} is for on-shell case. In this paper we want to extend this formalism to the  spin-3 topologically massive gravity. We find an expression for ADT conserved current which is off-shell conserved for a given Killing vector field. Then we can find the generalized ADT conserved charge by virtue of the Poincare lemma.
  In ref. \cite{31} the energy of the higher spin black hole solutions of ordinary higher spin gravity has obtained by canonical formalism.  Here we not only obtain the energy but also we find the angular momentum of black hole solutions of spin-3 topologically massive gravity. Furthermore we obtain a general formula for entropy of black hole solutions of spin-3 topologically massive gravity, where by substituting $e_{\mu} ^{\hspace{1.5 mm} ab}= \omega_{\mu} ^{\hspace{1.5 mm} ab}=0 $, this formula reduces to the entropy formula in the ordinary topologically massive gravity which we obtained previously in the paper \cite{9}.
\section{First-order formalism of 3D spin-3 gravity}
In this section, we will briefly review relevant aspects of spin-3 gravity in three dimensions in a slightly different way. We can describe the spin-3 gravity in three dimensions by the generalized dreibein and spin-connection \cite{1}. We want to work in the first order formalism, so we take generalized spin-connection as an independent quantity as well as generalized dreibein.\\
The algebra $sl(3, \mathbb{R})$ have 3 generators $J_{a}$ and 5 generators $T_{ab}$ with
the following commutation relations \footnote{In this paper, we use the ordinary symmetrization
 by a pair of parentheses, for instance $A_{(\mu} B_{\nu)} = \frac{1}{2} (A_{\mu} B_{\nu}+A_{\nu} B_{\mu})$, i.e.
we divide by the number of terms in the symmetrization.}
\begin{equation}\label{1}
  \begin{split}
      & [ J_{a} , J_{b} ] = \varepsilon _{ab} \hspace{0.1 mm} ^{c} J_{c}, \hspace{2 cm} [J_{a}, T_{bc}] = 2 \varepsilon ^{d} \hspace{0.1 mm} _{a ( b} T_{c) d}, \\
       & [T_{ab},T_{cd}] = -2 \left( \eta _{a(c} \varepsilon _{d)b} \hspace{0.1 mm} ^{e} + \eta _{b(c} \varepsilon _{d)a} \hspace{0.1 mm} ^{e} \right) J_{e} ,
  \end{split}
\end{equation}
where $T_{ab}$ are symmetric and trace-less in the Lorentz indices. In the representation we have considered the inner product of the generators are \cite{2}
\begin{equation}\label{2}
  \begin{split}
       & tr(J_{a}) =0 , \hspace{0.7 cm} tr(T_{ab}) =0, \hspace{0.7 cm}  tr(J_{a} T_{bc}) =0, \\
       & tr(J_{a} J_{b})=2 \eta _{ab}, \hspace{0.7 cm} tr(T_{ab} T_{cd}) = 4 \eta _{a(c} \eta _{d)b} - \frac{4}{3} \eta _{ab} \eta _{cd} .
  \end{split}
\end{equation}
The generalized dreibein and the generalized spin-connection take values in the Lie algebra $sl(3, \mathbb{R})$ as follows respectively:
\begin{equation}\label{3}
  e_{\mu} = e_{\mu} \hspace{0.1 mm} ^{a} J _{a} + e_{\mu} \hspace{0.1 mm} ^{ab} T _{ab},
\end{equation}
\begin{equation}\label{4}
  \omega _{\mu} = \omega _{\mu} \hspace{0.1 mm} ^{a} J _{a} + \omega _{\mu} \hspace{0.1 mm} ^{ab} T _{ab},
\end{equation}
 $a,b=0,1,2$. Notice that $e_{\mu} ^{\hspace{1.5 mm} ab} $ is symmetric and trace-less with respect to Lorentz indices. It is convenient to write the generalized dreibein one-form and the generalized spin-connection one-form as \cite{3,4}
\begin{equation}\label{5}
  e = e _{\mu} \hspace{0.1 mm} ^{A} J _{A} d x ^{\mu}, \hspace{1 cm} \omega = \omega _{\mu} \hspace{0.1 mm} ^{A} J _{A} d x ^{\mu},
\end{equation}
respectively, and $A=1,...,8$. In the above $J_{A}$ denotes a full set of $sl(3, \mathbb{R})$ generators,
\begin{equation}\label{6}
  [ J_{A} , J_{B} ] = f _{AB} \hspace{0.1 mm} ^{C} J_{C},
\end{equation}
where $f _{AB} \hspace{0.1 mm} ^{C}$ are anti-symmetric structure constants.
The Killing form in the fundamental representation of $sl(3, \mathbb{R})$ is defined as
\begin{equation}\label{7}
  K_{AB}=\frac{1}{2} tr(J_{A} J_{B}),
\end{equation}
and the anti-symmetric and symmetric structure constants of the Lie algebra are given by
\begin{equation}\label{8}
  f_{ABC}=\frac{1}{2} tr([ J_{A} , J_{B} ]J_{C}), \hspace{1 cm} d_{ABC}=\frac{1}{2} tr(\{ J_{A} , J_{B} \} J_{C}) .
\end{equation}
Translating the frame-like formalism to the metric-like formalism given by the following transformations \cite{1}
\begin{equation}\label{9}
\begin{split}
     & g_{\mu \nu}=\frac{1}{2!} tr(e_{(\mu} e_{\nu )}) \\
     & \varphi _{\mu \nu \lambda} = \frac{1}{3!} tr(e_{(\mu} e_{\nu} e_{\lambda )}).
\end{split}
\end{equation}
where $g_{\mu \nu}$ is the metric and $\varphi _{\mu \nu \lambda}$ is the spin-3 field and it is totally symmetric. By substituting $e_{\mu}$ from Eq.\eqref{5} into  Eq.\eqref{9} and using Eq.\eqref{7} and Eq.\eqref{8} we have
\begin{equation}\label{10}
\begin{split}
     & g_{\mu \nu}= K_{AB} e _{\mu} \hspace{0.1 mm} ^{A} e _{\nu} \hspace{0.1 mm} ^{B}\\
     & \varphi _{\mu \nu \lambda} = \frac{1}{3!} d_{ABC} e _{\mu} \hspace{0.1 mm} ^{A} e _{\nu} \hspace{0.1 mm} ^{B}e _{\lambda} \hspace{0.1 mm} ^{C}.
\end{split}
\end{equation}
It is clear that $g_{\mu \nu}$ and $\varphi _{\mu \nu \lambda}$ both are invariant under the following Lorentz-like gauge transformation
\begin{equation}\label{11}
  \tilde{e}_{\mu} = L e _{\mu} L^{-1},
\end{equation}
where $ L \in SL(3, \mathbb{R})$ and we can write $L = \exp{ (\lambda)}$ so that $\lambda$ is the generator of Lorentz-like transformation and it is a $sl(3, \mathbb{R})$ Lie algebra valued quantity,
\begin{equation}\label{11'}
 \lambda=\lambda^{a}J_{a}+\lambda^{ab}T_{ab}=\lambda^{A}J_{A}.
\end{equation}
Now we can introduce the exterior covariant derivative by demanding that it be covariant under the Lorentz-like transformation \eqref{11} as well as diffeomorphism covariance. To this end, we need to introduce the generalized spin-connection in the form of \eqref{5}. Thus, we can define exterior covariant derivative as
\begin{equation}\label{12}
  D_{\mu} e_{\nu} = \partial _{\mu} e_{\nu} + [ \omega _{\mu} , e _{\nu}] \hspace{0.5 cm} \Leftrightarrow
  \hspace{0.5 cm} D_{\mu} e _{\nu} \hspace{0.1 mm} ^{A} = \partial _{\mu} e  _{\nu} \hspace{0.1 mm} ^{A} + f ^{A} \hspace{0.1 mm} _{BC} \omega _{\mu} \hspace{0.1 mm} ^{B} e _{\nu} \hspace{0.1 mm} ^{C} .
\end{equation}
In order the above exterior derivative be covariant under the Lorentz-like transformation, the generalized spin-connection must transforms as
\begin{equation}\label{13}
  \tilde{ \omega } = L \omega L^{-1} +L d L ^{-1},
\end{equation}
where $d$ denotes the ordinary exterior derivative. Now, one can define the total derivative as follows:
\begin{equation}\label{14}
\begin{split}
   D ^{(T)} _{\mu} e_{\nu} = & \nabla _{\mu} e_{\nu} + [ \omega _{\mu} , e _{\nu}] \\
     =  & \partial _{\mu} e_{\nu} - \Gamma ^{\lambda} _{\mu \nu} e_{\lambda}+ [ \omega _{\mu} , e _{\nu}]
\end{split}
\end{equation}
where $ \Gamma ^{\lambda} _{\mu \nu} $ can be interprets as Affine connection and $\nabla$ denotes the ordinary
covariant derivative which is defined by using Affine connection. The metric-connection compatibility condition $\nabla _{\lambda} g_{\mu \nu} =0$ leads to the total derivative of dreibein vanishes, i.e. $D ^{(T)} _{\mu} e_{\nu}=0$. Therefore, one can define the generalized torsion 2-form as
\begin{equation}\label{15}
  \mathcal{T}= e_{\lambda} \Gamma ^{\lambda} _{\mu \nu} dx^{\mu} \wedge dx^{\nu} = D e.
\end{equation}
Another useful covariant quantity is the generalized curvature 2-form which is given by
\begin{equation}\label{16}
  \mathcal{R} = d \omega + \omega \wedge \omega \hspace{0.5 cm} \Leftrightarrow
  \hspace{0.5 cm} \mathcal{R} ^{A} = d \omega ^{A} + \frac{1}{2} f ^{A} \hspace{0.1 mm} _{BC} \omega ^{B} \wedge \omega ^{C} .
\end{equation}
Since the total derivative of $e_{\mu} \hspace{0.1 mm} ^{A}$ is zero, then we have $ \nabla _{\sigma} \varphi _{\mu \nu \lambda} =0 $.\\
From Eq.\eqref{11} and Eq.\eqref{13}, it is easy to check that the variation of generalized dreibein and generalized spin connection under an infinitesimal Lorentz-like gauge transformation are
\begin{equation}\label{17}
  \delta _{\lambda} e = [\lambda , e],
\end{equation}
\begin{equation}\label{18}
  \delta _{\lambda} \omega = - D \lambda ,
\end{equation}
where
\begin{equation}\label{18'}
  D_{\mu}=\partial_{\mu}\lambda+[\omega_{\mu}, \lambda].
\end{equation}
It is well-known that the ordinary Lie derivative of a diffeomorphism invariant quantity is diffeomorphism covariant but it may be not Lorentz-like covariant. In the manner of papers \cite{5,6,7,8,9}, we can generalize the Lie derivative so that it becomes covariant under Lorentz-like transformation as well as diffeomorphism. To this end, we consider the ordinary Lie derivative of dreibein along a curve generated by the vector filed $\xi$, $ \pounds _{\xi} e_{\mu}=\xi^{\lambda}\partial_{\lambda}e_{\mu}+e_{\lambda}\partial_{\mu}\xi^{\lambda}$. Now we generalize this by adding variation of $e_{\mu}$ with respect to an infinitesimal Lorentz-like gauge transformation
\begin{equation}\label{19}
  \mathfrak{L}_{\xi} e = \pounds _{\xi} e + [\lambda _{\xi} , e] .
\end{equation}
If the generator of Lorentz-like gauge transformation $\lambda _{\xi}$ transforms as
\begin{equation}\label{20}
  \tilde{\lambda} _{\xi} = L \lambda _{\xi} L^{-1} + L \pounds _{\xi} L ^{-1},
\end{equation}
under Lorentz-gauge transformation, then the definition of generalized Lie derivative \eqref{19} will be covariant
 under the Lorentz-like transformations. It is straightforward to extend this expression of the
generalized Lie derivative for $e _{\mu} \hspace{0.1 mm} ^{A}$ to the case for which we
 have more than one Lorentz-like index.
We demand that $\lambda _{\xi}$ depends on the vector field $\xi$ generating the diffeomorphism
and it is not depends on fields in general but in section 5, we will fix it such that Lie derivative
 of $e_{\mu}$ along $\xi$ vanishes explicitly when $\xi$ is a Killing vector field. The generalized Lie derivative of generalized spin-connection, $\mathfrak{L}_{\xi} \omega $, is not covariant under Lorentz-like transformations, while $\mathfrak{L}_{\xi} \omega - d \lambda _{\xi} $ will be covariant under the Lorentz-like transformations. Thus, under the generalized local translations we have
\begin{equation}\label{21}
  \delta _{\xi} e = \mathfrak{L}_{\xi} e,
\end{equation}
\begin{equation}\label{22}
  \delta _{\xi} \omega = \mathfrak{L}_{\xi} \omega - d \lambda _{\xi},
\end{equation}
 therefore they are covariant under the Lorentz-like transformations as well as diffeomorphism.
 Hence, we can interpret $\delta _{\xi}$ as a generalized diffeomorphism transformation.
We can use  Eq.\eqref{21} to obtain behaviour of the metric under this generalized diffeomorphism transformation as
\begin{equation}\label{23}
  \delta _{\xi} g_{\mu \nu} = \pounds _{\xi} g_{\mu \nu}.
\end{equation}
This equation is exactly what we expect. In a similar way, for spin-3 field we obtain
\begin{equation}\label{24}
  \delta _{\xi} \varphi _{\mu \nu \lambda} = \pounds _{\xi} \varphi _{\mu \nu \lambda},
\end{equation}
In the last step of the calculation we used $ f^{E}  _{ \hspace{1.5 mm} D (A } d_{BC) E} =0 $. Since $ \nabla _{\sigma} \varphi _{\mu \nu \lambda}$ is zero, the equation \eqref{24} can be rewritten as
\begin{equation}\label{25}
  \delta _{\xi} \varphi _{\mu \nu \lambda} = 3 \nabla _{( \lambda } \left(  \varphi _{\mu \nu ) \sigma} \xi ^{\sigma} \right) +
  3 \left( i_{\xi} \mathcal{T} ^{\sigma} \right)_{( \lambda} \varphi _{\mu \nu ) \sigma} ,
\end{equation}
thus in a torsion free theory this reduce to
\begin{equation}\label{26}
  \delta _{\xi} \varphi _{\mu \nu \lambda} = \nabla _{( \lambda } \xi _{\mu \nu )} ,
\end{equation}
where
\begin{equation}\label{27}
  \xi _{\mu \nu} = 3 \varphi _{\mu \nu  \sigma} \xi ^{\sigma} = 3 e_{\mu} ^{ \hspace{1.5 mm} a} e_{\nu} ^{ \hspace{1.5 mm} b} e_{\sigma ab} \xi ^{\sigma} - 4 e_{\mu} ^{ \hspace{1.5 mm} ab } e_{\nu bc} e_{\sigma \hspace{1.5 mm} a } ^{ \hspace{1.5 mm} c} \xi ^{\sigma} .
\end{equation}
The equation \eqref{26} is just the diffeomorphism part of the spin-3 field gauge transformation
which was introduced in \cite{10,11}, so it is clear from Eq.\eqref{26} that any transformation
under diffeomorphism can be written as a spin-3 field gauge transformation for a torsion free theory.\\
It is well known that Einstein-Hilbert action in the presence of negative cosmological constant in 3-dimension can be reformulated as a Chern-Simons theory with gage group $SO(2,2)\sim SL(2,\mathbb{R}) \times SL(2,\mathbb{R})$ \cite{w,a}. Similarly, a  $SL(3,\mathbb{R}) \times SL(3,\mathbb{R})$ Chern-Simons theory with the following action describes a three dimensional spin-3 gravity theory \cite{2',2''} (see also \cite{13}),
\begin{equation}\label{28}
  S_{EH}= S_{CS} [A ^{+}] - S_{CS}[A^{-}],
\end{equation}
where
\begin{equation}\label{29}
   S_{CS} [A] = \frac{l}{8 \pi G} \int ( A \wedge d A + \frac{2}{3} A \wedge A \wedge A ) ,
\end{equation}
where $l^2>0$ corresponds to a negative cosmological constant, and $G$ is Newton's constant.
To relate these to the first order formalism based on dreibein and spin-connection we introduce following two independent connection one-form $A^{+}$ and $A^{-}$,
\begin{equation}\label{30}
  A ^{\pm} = \omega \pm \frac{1}{l} e .
\end{equation}
By substituting $A ^{\pm}$ from Eq.\eqref{30} into Eq.\eqref{28} we obtain the following action which describes a three dimensional spin-3 gravity theory
\begin{equation}\label{31}
  S_{EH}= \frac{1}{16 \pi G} \int ( e \wedge \mathcal{R} +\frac{1}{3l^{2}} e \wedge e \wedge e ),
\end{equation}
In the next section we will extend the above Einstein-Hilbert spin-3 gravity action to the spin-3 topologically massive gravity theory. The field equations come from above action are
\begin{equation}\label{32}
  \mathcal{T}=0, \hspace{1.5 cm} \mathcal{R}+ \frac{1}{l^{2}} e \wedge e =0.
\end{equation}
 Since $i_{\xi} \omega$ transforms like \eqref{20} then one can choose $\lambda _{\xi} = i_{\xi} \omega $. Notice that this choice is not unique. Using the field equations and by choosing $\lambda _{\xi} = i_{\xi} \omega $, Eq.\eqref{17} and Eq.\eqref{18} reduce to
\begin{equation}\label{33}
  \delta _{\Xi} e = D \Xi , \hspace{1.5 cm} \delta _{\Xi} \omega = \frac{1}{l^{2}} [e , \Xi] ,
\end{equation}
where $\Xi = i_{\xi} e$. These results are exactly the generalized local translations which were already mentioned in \cite{1,3,4}.
\section{Spin-3 topologically massive gravity }
In this section, we consider an action describing the spin-3 field coupled to TMG. This generalization has already been done \cite{12,7'}. In that paper, the authors have studied this model at the linearized level. In this paper, we are interested to consider the Spin-3 Topologically Massive Gravity in non-linear level and we want to find a general formula for quasi-local conserved charges of the solutions of this model.\\
The Lagrangian of the spin-3 Topologically Massive Gravity is given by {\color{red} \cite{12}}
\begin{equation}\label{34}
  \mathcal{L}= tr \{ - \sigma e \wedge \mathcal{R} + \frac{\Lambda}{3} e \wedge e\wedge e + \frac{1}{2 \mu} \left(\omega d \omega + \frac{2}{3} \omega \wedge \omega \wedge \omega \right) + h \wedge \mathcal{T} \}.
\end{equation}
In the above Lagrangian $h$ is an auxiliary $sl(3,\mathbb{R})$ Lie algebra valued one-form field. Also $\mu$ is a mass parameter, $\sigma$ is a sign and $\Lambda$ denotes the cosmological parameter.\\
The general variation of the Lagrangian \eqref{34} is given by
\begin{equation}\label{35}
  \delta \mathcal{L} = tr(\delta \Phi \wedge E_{\Phi})+ d \Theta (\Phi , \delta \Phi).
\end{equation}
We use the symbol $\Phi $ to denote all of fields $e$, $\omega$ and $h$. Here $E_{\Phi}$ denotes the equation of motion associated with
field $\Phi $. Equations of motion are given as follows:
\begin{equation}\label{36}
  E_{e} = - \sigma \mathcal{R} + \Lambda e \wedge e + D h =0 ,
\end{equation}
\begin{equation}\label{37}
  E_{\omega} = - \sigma \mathcal{T} + \frac{1}{\mu} \mathcal{R} + e \wedge h + h \wedge e =0 ,
\end{equation}
\begin{equation}\label{38}
  E_{h} = \mathcal{T} =0 .
\end{equation}
Also, the surface term $\Theta (\Phi , \delta \Phi)$ is
\begin{equation}\label{39}
  \Theta (\Phi , \delta \Phi) = tr \left(- \sigma \delta \omega \wedge e + \frac{1}{2 \mu} \delta \omega \wedge \omega + \delta e \wedge h \right).
\end{equation}
If we take
\begin{equation}\label{40}
  l^{2}= - \frac{\Lambda}{\sigma} \hspace{0.7 cm} and \hspace{0.7 cm} h= \frac{1}{2 \mu l^{2}} e,
\end{equation}
then the equations of motion \eqref{36}-\eqref{38} reduce to
\begin{equation}\label{41}
  \mathcal{T}=0, \hspace{1.5 cm} \mathcal{R}+ \frac{1}{l^{2}} e \wedge e =0.
\end{equation}
These are exactly what we are mentioned already in the equation \eqref{32}. In this way, all the solutions of the spin-3 gravity (for instance, see \cite{2,13,14}) are solutions of the spin-3 topologically massive gravity.

\section{Quasi-local conserved charges}
The action \eqref{31} of spin-3 gravity in 3D is invariant under the generalized local Lorentz translations \eqref{21} and \eqref{22}. In other words, variation of the Lagrangian of spin-3 gravity with respect to the generalized local Lorentz translations is equals to the generalized Lie derivative of the Lagrangian, i.e. $ \delta _{\xi} \mathcal{L}_{EH} = \mathfrak{L}_{\xi} \mathcal{L}_{EH} $. But, this is not the case for the spin-3 topologically massive gravity. Consider the spin-3 topologically massive gravity \eqref{34}. This Lagrangian is not invariant under the generalized local Lorentz translations. In this model, variation of the Lagrangian due to the generalized local Lorentz translations becomes
\begin{equation}\label{42}
  \delta _{\xi} \mathcal{L} = \mathfrak{L}_{\xi} \mathcal{L} + d \Psi _{\xi},
\end{equation}
where $\Psi _{\xi}$ is given by
\begin{equation}\label{43}
   \Psi _{\xi} = \frac{1}{2 \mu} tr(d \lambda _{\xi} \wedge \omega) .
\end{equation}
In the calculation of the equation \eqref{42}, we use the fact that the generalized Lie derivative do not commute with the exterior derivative,
\begin{equation}\label{44}
  [d, \mathfrak{L}_{\xi} ] e = d \lambda _{\xi} \wedge e + e \wedge \lambda _{\xi}.
\end{equation}
Also, notice that $\lambda _{\xi}$ does not depend on the fields. In a similar way, one can show that the variation of surface term \eqref{39} due to the generalized local Lorentz translations differs form its generalized Lie derivative as
\begin{equation}\label{45}
  \delta _{\xi} \Theta (\Phi , \delta \Phi) = \mathfrak{L}_{\xi} \Theta (\Phi , \delta \Phi) +  \Pi _{\xi},
\end{equation}
where $\Pi _{\xi}$ is given by
\begin{equation}\label{46}
   \Pi _{\xi} = \frac{1}{2 \mu} tr(d \lambda _{\xi} \wedge \delta \omega) .
\end{equation}
Now we consider variation of the Lagrangian Eq.\eqref{34} due to a generalized local Lorentz translation,
\begin{equation}\label{47}
  \delta _{\xi} \mathcal{L} = tr(\delta _{\xi} \Phi \wedge E_{\Phi})+ d \Theta (\Phi , \delta _{\xi} \Phi).
\end{equation}
By using Eqs.\eqref{42}, \eqref{47} and from $ \mathfrak{L}_{\xi} \mathcal{L} = \pounds _{\xi} \mathcal{L} = d i_{\xi} \mathcal{L} $, we can write
\begin{equation}\label{48}
  d \left( \Theta (\Phi , \delta _{\xi} \Phi) - i_{\xi} \mathcal{L} - \Psi _{\xi} \right) = - tr(\delta _{\xi} \Phi \wedge E_{\Phi}).
\end{equation}
We will not restrict our study to the on-shell case. One can rewrite Eq.\eqref{21} as follows:
\begin{equation}\label{49}
  \delta _{\xi} e = \pounds _{\xi} e+[\lambda _{\xi}, e]=\pounds _{\xi}+[i_{\xi}\omega, e]+[\lambda-i_{\xi}\omega, e] = D (i_{\xi} e) + i_{\xi} \mathcal{T} + [\lambda _{\xi} - i_{\xi} \omega , e],
\end{equation}
where $i_{\xi}\mathcal{T}=(i_{\xi}\mathcal{T}^{\sigma})_{\nu}e_{\sigma}dx^{\nu}$. Now we rewrite Eq.\eqref{22} as
\begin{equation}\label{50}
  \delta _{\xi} \omega =\pounds _{\xi}\omega +[\lambda _{\xi}, \omega]-d\lambda = i_{\xi} \mathcal{R} - D ( \lambda _{\xi}- i_{\xi} \omega).
\end{equation}
Also, we can show that
\begin{equation}\label{51}
  D ^{2} h =dDh+[\omega, Dh] =\mathcal{R} \wedge h - h \wedge \mathcal{R}.
\end{equation}
By some calculations and using Eqs.\eqref{49}-\eqref{51}, we can simplify the right hand side of the equation \eqref{48} as
\begin{equation}\label{52}
  \begin{split}
      tr(\delta _{\xi} \Phi \wedge E_{\Phi})= & d \hspace{1 mm} tr \left( i_{\xi} e E_{e} - ( \lambda _{\xi}- i_{\xi} \omega) E _{\omega} + i_{\xi} h E_{h} \right) \\
       & + tr \{ \sigma i_{\xi} e D \mathcal{R} - i_{\xi} h (D \mathcal{T} + e \wedge \mathcal{R} - \mathcal{R} \wedge e) \} \\
       & + tr \{ ( \lambda _{\xi}- i_{\xi} \omega) [ - \sigma (D \mathcal{T} + e \wedge \mathcal{R} - \mathcal{R} \wedge e) + \frac{1}{\mu} D \mathcal{R} ] \}.
  \end{split}
\end{equation}
By substituting Eq.\eqref{52} into Eq.\eqref{48} we will have
\begin{equation}\label{53}
\begin{split}
   d \mathfrak{J} _{\xi} = & tr \{ - \sigma i_{\xi} e D \mathcal{R} + i_{\xi} h (D \mathcal{T} + e \wedge \mathcal{R} - \mathcal{R} \wedge e) \} \\
     & + tr \{ ( \lambda _{\xi}- i_{\xi} \omega) [  \sigma (D \mathcal{T} + e \wedge \mathcal{R} - \mathcal{R} \wedge e) - \frac{1}{\mu} D \mathcal{R} ] \},
\end{split}
\end{equation}
where $\mathfrak{J} _{\xi}$ is given by
\begin{equation}\label{54}
  \mathfrak{J} _{\xi} = \Theta (\Phi , \delta _{\xi} \Phi) - i_{\xi} \mathcal{L} - \Psi _{\xi} + tr( i_{\xi} e E_{e} - ( \lambda _{\xi}- i_{\xi} \omega) E _{\omega} + i_{\xi} h E_{h} ).
\end{equation}
By virtue the following generalized Bianchi identities \footnote{ We can read off the Bianchi identities from
the action of considered theory of gravity. To clear this matter, consider the action of a gravity theory in
the metric formalism, $S_{gravity}= \int \mathcal{L}$. Under an infinitesimal diffeomorphism generated by a
 vector field
$\xi$, we have $\delta _{\xi} S_{gravity} = - \int \sqrt{-g}\mathcal{G}^{\mu \nu} \delta _{\xi} g_{\mu \nu} + $
surface term,
where $\mathcal{G}^{\mu \nu}$ is generalized Einstein tensor. Because $\delta _{\xi} g_{\mu \nu} = \nabla _{\mu} \xi _{\nu} + \nabla _{\nu} \xi _{\mu} $ then $\delta _{\xi} S_{gravity} = - 2 \int \sqrt{-g}\mathcal{G}^{\mu \nu} \nabla _{\mu} \xi _{\nu} +$
 surface term or $\delta _{\xi} S_{gravity} = 2 \int \sqrt{-g} \nabla _{\mu} \mathcal{G}^{\mu \nu} \xi _{\nu} +$
surface term. Dropping surface term by some boundary conditions, we see that the action is diffeomorphism
invariant if $\nabla _{\mu} \mathcal{G}^{\mu \nu} = 0$, for any non-zero vector field $\xi$, and this is
 exactly the Bianchi identities. Thus, using the invariant property of the action under diffeomorphism,
 we can read off the Bianchi identities. Here, we use the first order and off-shell version of this prescription
to read off the Bianchi identities in considered formalism}.
  \begin{equation}\label{55}
\begin{split}
     & D \mathcal{R} = 0, \\
     & D \mathcal{T} + e \wedge \mathcal{R} - \mathcal{R} \wedge e =0.
\end{split}
\end{equation}
the current $\mathfrak{J} _{\xi}$  is conserved off-shell, for arbitrary $\xi$,
because the right hand
side of Eq.\eqref{53} is zero, so $d \mathfrak{J} _{\xi} =0$. Thus, by the Poincare lemma we find that
\begin{equation}\label{56}
   \mathfrak{J} _{\xi} =d K _{\xi} .
\end{equation}
By varying the equation Eq.\eqref{54} and using Eq.\eqref{34} and Eq.\eqref{45}, we will have
\begin{equation}\label{57}
\begin{split}
     & d \left( \delta K _{\xi} - i_{\xi} \Theta (\Phi , \delta _{\xi} \Phi) \right) - \left( \delta \Theta (\Phi , \delta _{\xi} \Phi) - \delta _{\xi} \Theta (\Phi , \delta \Phi) \right) + \delta \psi _{\xi} - \Pi _{\xi} = \\
     & tr \{ \delta e \wedge i _{\xi} E _{e} + \delta \omega \wedge i _{\xi} E _{\omega} + \delta h \wedge i _{\xi} E _{h} + i _{\xi} e \delta E _{e} + i _{\xi} h \delta E _{h} - ( \lambda _{\xi}- i_{\xi} \omega) \delta E _{\omega} \} \\
\end{split}
\end{equation}
Now, we restrict our attention to the case in which $\xi$ is a Killing vector field.
On the one hand, we have the following configuration space result given in \cite{15}
\begin{equation}\label{58}
  \delta \Theta (\Phi , \delta _{\xi} \Phi) - \delta _{\xi} \Theta (\Phi , \delta \Phi)=0,
\end{equation}
this equality holds when $\xi$ is a Killing vector field. On the other hand, the right hand side of the equation \eqref{57} is just the first order formalism version of the off-shell ADT conserved current for the considered theory
\begin{equation}\label{59}
  \mathcal{J} _{ADT} = tr \{ \delta e \wedge i _{\xi} E _{e} + \delta \omega \wedge i _{\xi} E _{\omega} + \delta h \wedge i _{\xi} E _{h} + i _{\xi} e \delta E _{e} + i _{\xi} h \delta E _{h} - ( \lambda _{\xi}- i_{\xi} \omega) \delta E _{\omega} \}.
\end{equation}
One can find the metric formalism version of the off-shell ADT conserved current in \cite{16,17}. The definition $\mathcal{J} _{ADT}$ by Eq.\eqref{59} is just resemble of the metric formalism version. Thus, we can rewrite Eq.\eqref{57} as follows:
\begin{equation}\label{60}
   \mathcal{J} _{ADT} = d \left( \delta K _{\xi} - i_{\xi} \Theta (\Phi , \delta _{\xi} \Phi) \right).
\end{equation}
In the last step of the calculations we used of the fact that $\delta \Phi _{\xi} - \Pi _{\xi} =0$ which can be easily read off from Eq.\eqref{43} and Eq.\eqref{46}. Now, it is obvious that $\mathcal{J} _{ADT}$ is conserved off-shell for any Killing vector field $\xi$ and we can define the off-shell ADT conserved charge associated with a Killing vector field $\xi$ as
\begin{equation}\label{61}
  \mathcal{Q} _{ADT} (\Phi , \delta \Phi ; \xi) = \delta K _{\xi} - i_{\xi} \Theta (\Phi , \delta _{\xi} \Phi),
\end{equation}
so that through it we can rewrite Eq.\eqref{60} as $ \mathcal{J} _{ADT} = d \mathcal{Q} _{ADT} $. Hence, in the manner of papers \cite{16,17}, we define the quasi-local charge corresponding to a Killing vector $\xi$ as
\begin{equation}\label{62}
  Q(\xi) =\frac{1}{8 \pi } \int_{0}^{1} ds \int _{\Sigma} \mathcal{Q} _{ADT} (\Phi | s) ,
\end{equation}
where $\Sigma$ denotes a codimension two space-like surface and integration with respect
to $s$ runs over a one-parameter path in the solution space. Also, $s = 0$ and $s = 1$ correspond to the background solution and the interested
solution, respectively.\\
By substituting the explicit expressions of the surface term Eq.\eqref{39}, the Lagrangian Eq.\eqref{34} and Eq.\eqref{43} into Eq.\eqref{54} we obtain $\mathfrak{J} _{\xi} $, then using Eq.\eqref{56}, we find that
\begin{equation}\label{63}
  K _{\xi} = tr \{ \sigma (\lambda _{\xi} - i _{\xi} \omega ) e + i _{\xi} e h + \frac{1}{2 \mu} i _{\xi} \omega \omega - \frac{1}{\mu} \lambda _{\xi} \omega \}.
\end{equation}
This ensures that $ \mathfrak{J} _{\xi} $ is closed and exact off-shell and hence the equations in \eqref{55} must be satisfied. Hence, we can express the ADT conserved charge \eqref{61} explicitly as follows:
\begin{equation}\label{64}
  \mathcal{Q} _{ADT} (\Phi , \delta \Phi ; \xi) = tr \{ \sigma (\lambda _{\xi} - i _{\xi} \omega ) \delta e + i _{\xi} e \delta h -
  \frac{1}{\mu} (\lambda _{\xi} - i _{\xi} \omega ) \delta \omega + i _{\xi} h \delta e - \sigma i_{\xi} e \delta \omega \} .
\end{equation}
In the section 3, we showed that all the solutions of spin-3 gravity in three dimensions are solutions of the spin-3 topologically massive gravity in the presence of following relations
\begin{equation}\label{67}
  \Lambda = - \frac{\sigma}{l^{2}} \hspace{0.7 cm} and \hspace{0.7 cm} h= \frac{1}{2 \mu l^{2}} e.
\end{equation}
By substituting the above $h$ into Eq.\eqref{64} we find that
\begin{equation}\label{68}
  \mathcal{Q} _{ADT} (\Phi , \delta \Phi ; \xi) = tr \{ (\lambda _{\xi} - i _{\xi} \omega ) ( \sigma \delta e - \frac{1}{\mu} \delta \omega ) + i _{\xi} e ( - \sigma \delta \omega + \frac{1}{\mu l^{2}} \delta e )  \} .
\end{equation}
It is traditional to write conserved charge in terms of gauge connections $A^{\pm}$. Using Eq.\eqref{30} we have
 \begin{equation}\label{68'}
  e=\frac{l}{2}( A ^{+}-A ^{-}), \hspace{1cm} \omega=\frac{1}{2}( A ^{+}+A ^{-}).
\end{equation}
Then we can rewrite the off-sell ADT conserved charge corresponds to a Killing vector field as follows:
\begin{equation}\label{69}
  \mathcal{Q} _{ADT} (\Phi , \delta \Phi ; \xi) = \frac{l}{2} tr \{ (\sigma - \frac{1}{\mu l})( \lambda _{\xi} - i _{\xi} A ^{+} ) \delta A ^{+} -
  (\sigma + \frac{1}{\mu l})( \lambda _{\xi} - i _{\xi} A ^{-} ) \delta A ^{-} \} .
\end{equation}
If we substitute $\lambda _{\xi}=0$ in Eq.\eqref{69}, in another word if we consider only covariance under diffeomorphism, then ADT conserved charge reduce to the following
\begin{equation}\label{71}
  \mathcal{Q} _{ADT} (\Phi , \delta \Phi ; \xi) = \frac{-l}{2} tr \{ (\sigma - \frac{1}{\mu l}) i _{\xi} A ^{+}  \delta A ^{+} -
  (\sigma + \frac{1}{\mu l})  i _{\xi} A ^{-} \delta A ^{-} \} .
\end{equation}
Now if we consider the surface $\sum$ in Eq.\eqref{62} as a circle with infinite radius, and insert the above $ \mathcal{Q} _{ADT}$ into  Eq.\eqref{62} we obtain

\begin{equation}\label{72}
  Q(\xi) = - \frac{l}{16 \pi } \lim_{\rho\rightarrow\infty}\int_{0}^{2\pi}  tr \{ (\sigma - \frac{1}{\mu l}) i _{\xi} A ^{+}  \delta A _{\phi} ^{+} -
  (\sigma + \frac{1}{\mu l})  i _{\xi} A ^{-} \delta A _{\phi} ^{-} \} d\phi .
\end{equation}
If we take the limit $\frac{1}{\mu}\rightarrow 0$, where TMG action Eq.\eqref{34} reduce to the Einstein-Hilbert action Eq.\eqref{31}, the above conserved charge Eq.\eqref{72} also in this limiting case reduce to the result presented in \cite{31} exactly (see also \cite{32}). So we could obtain an extended version of conserved charge presented in \cite{31} in the first order formalism and in off-shell case by the quasi-local approach.
\section{Gauge fixing}
In this section, we want to find an expression for $\lambda _{\xi}$ when the generalized Lie derivative of generalized dreibein Eq. \eqref{19} vanishes explicitly where $\xi$ is a Killing vector field. We suggest following expression
\begin{equation}\label{73}
  \lambda _{\xi}= i _{\xi} \omega + \frac{1}{2} [e ^{\nu} , D_{\nu} (i_{\xi} e) +(i_{\xi}\mathcal{T})_{\nu} ].
\end{equation}
The above $\lambda _{\xi}$ manifestly transforms like Eq.\eqref{20} under the Lorentz-like gauge transformation. In the considered Lie algebra, the following relation holds between Killing form and the anti-symmetric structure constants
\begin{equation}\label{74}
  f _{AB} ^{\hspace{4 mm} E} f_{E C D} = - (K _{AC} K_{BD} - K _{AD} K_{BC} ).
\end{equation}
By substituting Eq.\eqref{73} into Eq.\eqref{19}, and using the metric-connection compatibility condition and the above equation we find that
\begin{equation}\label{75}
  \mathfrak{L}_{\xi} e _{\mu}= \frac{1}{2} e^{\nu} \pounds _{\xi} g_{\mu \nu} .
\end{equation}
It is clear now that $\mathfrak{L}_{\xi} e $ becomes zero when $\xi$ is a Killing vector field, where we have $\pounds _{\xi} g_{\mu \nu}=0$.\\
We have introduced $\lambda$ to be the generator of Lorentz-like gauge transformation (see Eq.\eqref{11}). In the equation \eqref{11}, $\lambda$ is just an arbitrary function of coordinates. It is easy to see that the ordinary Lie derivative of a Lorentz-like invariant quantity, say $e$, is not Lorentz-like covariant. On the other hand, the change of $e$ under Lorentz-like gauge transformation is given by Eq.\eqref{17}. It was shown that by combining the change due to infinitesimal Lorentz-like gauge transformation with ordinary Lie derivative one can define generalized Lie derivative \eqref{19} which is Lorentz-like covariant provided that $\lambda$ transforms as \eqref{20} under Lorentz-like gauge transformation. Simply one can see form Eq.\eqref{19} that the change of $\lambda$ , under infinitesimal Lorentz-like gauge transformation, is given by $\delta \lambda = - \pounds _{\xi} \lambda$. So, we expect that $\lambda$ to be also a function of $\xi$, i.e. $\lambda =  \lambda _{\xi} (x)$. Another reason for this comes from the fact that we set total variation due to $\xi$ equal to generalized Lie derivative with respect to $\xi$ and $\lambda _{\xi}$ is obliged to generate variation due to $\xi$ (see Eq.\eqref{21}). Thus, in the formalism presented in this paper, $\lambda$ should be a function of coordinates and of $\xi$. We need to have an expression for $\lambda _{\xi}$ in terms of dynamical fields when we calculate conserved charges of a solution. To this end, we demanded that the generalized Lie derivative of $e$ explicitly vanishes when $\xi$ is a Killing vector field (see Eq.\eqref{75}). Hence, to obtain a general expression for the conserved charge \eqref{62} corresponds to a Killing vector field we must consider $\lambda _{\xi}$ as a function of coordinates and of $\xi$. Eventually, to calculate conserved charges of a solution (by using Eq.\eqref{62}) we need to have an expression for $\lambda _{\xi}$ in terms of dynamical fields so, we substitute $\lambda _{\xi}$ from Eq.\eqref{73}.
\section{A general formula for the entropy of black holes}
Let us consider a black hole solution of the spin-3 topologically massive gravity.
From section 4 we know that any quasi-local conserved charge of black hole corresponding to the Killing vector
 field $\xi$ is given by Eq.\eqref{62}. Robert. Wald has proposed that entropy of black holes is related to
the concept of conserved charge \cite{41}. He show that entropy of a black hole is proportional to the conserved
charge associated to the Killing vector field generating Killing horizon. Here, we follow the method that proposed by
 Wald which provides a general formula for entropy of black hole. We take the codimension two surface $\Sigma$ to be the bifurcate surface $\mathcal{B} $. Suppose that $\xi$ is the Killing vector field which generates the Killing horizon, so we must set $\xi=0$ on $\mathcal{B} $. Thus, by substituting $\mathcal{Q} _{ADT}$  from Eq.\eqref{64} into Eq.\eqref{62}, the conserved charge expression \eqref{62} reduces to
\begin{equation}\label{76}
  Q(\xi) =\frac{1}{8 \pi } \int_{0}^{1} ds \int _{\mathcal{B}} tr \{ \lambda _{\xi} (\sigma \delta e  - \frac{1}{\mu} \delta \omega ) \}.
\end{equation}
 Now, we take $s=0$ and $s=1$ correspond to the considered black hole space-time an the perturbed one, respectively. Therefore, the equation \eqref{76} becomes
\begin{equation}\label{77}
  \delta Q(\xi) =\frac{1}{8 \pi } \int _{\mathcal{B}} tr \{ \lambda _{\xi} (\sigma \delta e  - \frac{1}{\mu} \delta \omega ) \} .
\end{equation}
As we mentioned earlier, $\lambda _{\xi}$ do not depends on the fields at all, so we have
\begin{equation}\label{78}
   Q(\xi) =\frac{1}{8 \pi } \int _{\mathcal{B}} tr \{ \lambda _{\xi} (\sigma  e  - \frac{1}{\mu} \omega ) \} .
\end{equation}
On the bifurcate surface \footnote{We assume that the considered theory may be has some black hole solutions (see also our discussion in footnote 6)
 and we know that a black hole has an outer event horizon on which the norm of Killing vector field
 generating event horizon vanishes. Consider a general stationary metric as \eqref{98}.
 The considered spacetime describes a black hole when it has event horizon.
If the equation $g_{t \phi} ^{2}- g_{tt} g_{\phi \phi}=0$ has a real solution for radial coordinate
 $\rho=\rho _{H}$ then the considered spacetime has an event horizon. We know that,
 $\xi ^{\mu} \xi _{\mu} =0$ at $\rho=\rho _{H}$ for Killing vector field generating Killing horizon and
 this happens if $\rho _{H}$ exists as a solution of the equation
 $g_{t \phi} ^{2}- g_{tt} g_{\phi \phi}=0$ \cite{42}.
 So, we can take $\Sigma$ as a codimension two space-like surface to be bifurcate surface.
 We will find a suitable $\xi$ for an example in section 7. In the other hand, an important property of higher-spin gravity is that the
  gauge transformation of higher-spin field act nontrivially on the metric. Due to this feature of the theory, the event horizon of black hole become gauge dependent \cite{34'}. The authors of \cite{34'} have discussed on the solutions of spin-3 gravity previously introduced in \cite{13}, and have shown that those solutions describe a traversable wormhole connecting two asymptotic region, instead a black hole. Then they have shown that under a higher spin gauge transformation these solutions can be transformed
to describe black holes with manifestly smooth event horizons. }, we can rewrite Eq.\eqref{73} as
\begin{equation}\label{79}
  \lambda _{\xi}= \frac{1}{2} \nabla ^{\mu} \xi ^{\nu} [e _{\mu} , e_{\nu}].
\end{equation}
Since we have $\nabla ^{\mu} \xi ^{\nu} = \kappa n^{\mu \nu}$ on the bifurcate surface, where $\kappa$ is surface gravity and $n^{\mu \nu}$ is bi-normal to $\mathcal{B} $, therefore Eq.\eqref{79} becomes
\begin{equation}\label{80}
  \lambda _{\xi}= \frac{1}{2} \kappa n^{\mu \nu} [e _{\mu} , e_{\nu}] = \frac{1}{2} \kappa f^{A}_{\hspace{1.5 mm} BC} e _{\mu}^{\hspace{1.5 mm} B}  e_{\nu} ^{\hspace{1.5 mm} C} n^{\mu \nu} J_{A} .
\end{equation}
Now, we define $N^{A}$ as follows:
\begin{equation}\label{81}
  N^{A} = \frac{1}{2} f^{A}_{\hspace{1.5 mm} BC} e _{\mu}^{\hspace{1.5 mm} B}  e_{\nu} ^{\hspace{1.5 mm} C} n^{\mu \nu},
\end{equation}
since the bi-normal to $\mathcal{B} $ is normalized to $-2$ then $N^{A}$ is normalized to $1$, i.e. $ K_{AB} N^{A} N^{B}=1$. For the stationary black hole solutions, the non-zero components of the bi-normal to $\mathcal{B} $ are $n^{t \rho} = - n^{\rho t}$, therefore Eq.\eqref{81} becomes
\begin{equation}\label{82}
  N^{A} = f^{A}_{\hspace{1.5 mm} BC} e _{t}^{\hspace{1.5 mm} B}  e_{\rho} ^{\hspace{1.5 mm} C} n^{t \rho},
\end{equation}
where $t$ and $\rho$ are time-coordinate and radial-coordinate, respectively. It is obvious that $N^{A}$ is orthogonal to $e _{t}^{\hspace{1.5 mm} A} $ and $ e_{\rho} ^{\hspace{1.5 mm} A}$, i.e. $ K_{AB} e _{t}^{\hspace{1.5 mm} A} N^{B} = K_{AB} e _{\rho}^{\hspace{1.5 mm} A} N^{B} = 0$,  then it must be proportional to $ e _{\phi}^{\hspace{1.5 mm} A}$, where $\phi$ denotes the angular coordinate. Because $N^{A}$ is normalized to $1$ then we have
\begin{equation}\label{83}
  N^{A} = \frac{1}{\sqrt{ g_{\phi \phi}}} e _{\phi}^{\hspace{1.5 mm} A},
\end{equation}
by substituting this equation into Eq.\eqref{80} we find that
\begin{equation}\label{84}
  \lambda_{\xi} = \frac{\kappa}{\sqrt{ g_{\phi \phi}}} e _{\phi}.
\end{equation}
We substitute the above expression into Eq.\eqref{78} then we obtain
\begin{equation}\label{85}
   Q(\xi) =\frac{\kappa}{8 \pi } \int _{\mathcal{B}} \frac{1}{\sqrt{ g_{\phi \phi}}} tr \{ e _{\phi} (\sigma  e _{\phi} - \frac{1}{\mu} \omega _{\phi} ) \} d \phi .
\end{equation}
Now, we can define the entropy of a black hole as
\begin{equation}\label{86}
   S= \frac{8 \pi}{\kappa} Q(\xi) = \int _{\mathcal{B}} \frac{1}{\sqrt{ g_{\phi \phi}}} tr \{ e _{\phi} (\sigma  e _{\phi} - \frac{1}{\mu} \omega _{\phi} ) \} d \phi .
\end{equation}
If we set $e_{\mu} ^{\hspace{1.5 mm} ab}= \omega_{\mu} ^{\hspace{1.5 mm} ab}=0 $, then the above formula reduces to the entropy formula in the ordinary topologically massive gravity which is obtained in the paper \cite{9}.
\section{Example}
 The authors of  \cite{13} (see also \cite{1}) have proposed following gauge connections which solve the spin-3 gravity field equations
\begin{equation}\label{87}
  \begin{split}
       & A^{+} = b(\rho)^{-1} a^{+}(x^{+}) b(\rho)+b(\rho)^{-1} d b(\rho) \\
       & A^{-} = b(\rho) a^{-}(x^{-}) b(\rho)^{-1}+b(\rho) d b(\rho)^{-1},
  \end{split}
\end{equation}
where $b(\rho)= \exp{(\rho L_{0})}$ and $x^{\pm}= t/l \pm \phi$. Also, $a^{\pm}(x^{\pm})$ are given by
\begin{equation}\label{88}
  \begin{split}
       & a^{+}(x^{+})= \left( L_{1} - \mathcal{L}^{+}(x^{+}) L_{-1} - \mathcal{W}^{+}(x^{+}) W_{-2} \right) dx^{+} \\
       & a^{-}(x^{-})= \left( L_{-1} - \mathcal{L}^{-}(x^{-}) L_{1} + \mathcal{W}^{-}(x^{-}) W_{2} \right) dx^{-}.
  \end{split}
\end{equation}
where $\mathcal{L}^{\pm}(x^{\pm})$ and $\mathcal{W}^{\pm}(x^{\pm})$ are functions which transform under
 gauge transformations which preserve the asymptotic conditions (see \cite{1} for details). A representation of
 generators $L_{i}, (i=-1,0,1)$, $W_{m}, (m=-2,-1,0,1,2)$ is given in appendix. By substituting the gauge connections \eqref{87} into Eq.\eqref{68'} we can find the generalized dreibein as
\begin{equation}\label{89}
  \begin{split}
       & e_{t}=\frac{1}{2} \left( (e^{\rho} - \mathcal{L}^{-} e^{-\rho} ) L_{1} + ( e^{\rho} - \mathcal{L}^{+} e^{-\rho} ) L_{-1} + \mathcal{W}^{-} e^{-2\rho} W_{2} - \mathcal{W}^{+} e^{-2\rho} W_{-2} \right) \\
       & e_{\phi}=\frac{l}{2} \left( (e^{\rho} + \mathcal{L}^{-} e^{-\rho} ) L_{1} - ( e^{\rho} + \mathcal{L}^{+} e^{-\rho} ) L_{-1} - \mathcal{W}^{-} e^{-2\rho} W_{2} - \mathcal{W}^{+} e^{-2\rho} W_{-2} \right) \\
       & e_{\rho}= l L_{0}.
  \end{split}
\end{equation}
Similarly, the space-time components of the generalized spin-connections can be find as
\begin{equation}\label{90}
  \omega _{t} = \frac{1}{l^2} e_{\phi}, \hspace{0.7 cm} \omega _{\rho} =0, \hspace{0.7 cm} \omega _{\phi} = e _{t}.
\end{equation}
By substituting generalized dreibeins  Eq.\eqref{89} into Eq.\eqref{9}, the metric takes following form
\begin{equation}\label{91}
  \begin{split}
     ds^{2} = & - \left( \mathcal{L}^{+} \mathcal{L}^{-} e^{-2 \rho} + 4 \mathcal{W}^{+} \mathcal{W}^{-} e^{-4 \rho} - \mathcal{L}^{+} - \mathcal{L}^{-} +e^{2 \rho} \right) dt^{2} \\
     & + l^{2} d \rho ^{2} + 2 l (\mathcal{L}^{+} - \mathcal{L}^{-}) dt d \phi \\
       & + l^{2} \left( \mathcal{L}^{+} \mathcal{L}^{-} e^{-2 \rho} + 4 \mathcal{W}^{+} \mathcal{W}^{-} e^{-4 \rho} + \mathcal{L}^{+} + \mathcal{L}^{-} +e^{2 \rho} \right) d \phi ^{2}.
  \end{split}
\end{equation}
Now, we try to find energy and angular momentum of this solution. To this end, we take co-dimension two surface to be a circle at infinity and the background solution corresponds to the case in which $\mathcal{L}^{\pm}=\mathcal{W}^{\pm}=0$. So, for the background solution, we have
\begin{equation}\label{92}
  \begin{split}
       & \bar{e} _{t}=\frac{1}{2} e^{\rho} \left( L_{1} + L_{-1}  \right) \\
       & \bar{e} _{\phi}=\frac{l}{2} e^{\rho} \left(  L_{1} -  L_{-1} \right) \\
       & \bar{e} _{\rho}= l L_{0} .
  \end{split}
\end{equation}
The bar on the top of a quantity emphasizes that quantity calculated on background. Because the spin-3 topologically massive gravity is a torsion-free theory, Eq.\eqref{73} reduce to
\begin{equation}\label{93}
  \lambda _{\xi} - i _{\xi} \omega = \nabla ^{\mu} \xi ^{\nu} e_{\mu} e_{\nu} .
\end{equation}
Energy corresponds to the Killing vector $\xi_{(t)}= - \partial _{t} $. For this Killing vector on the background, Eq.\eqref{93} becomes
\begin{equation}\label{94}
  \bar{\lambda} _{\xi _{(t)}} - i _{\xi _{(t)}} \bar{\omega} = \frac{1}{l^{2}} \bar{e}_{\phi}.
\end{equation}
By substituting Eqs.\eqref{89}, \eqref{90}, \eqref{92} and \eqref{94} into Eq.\eqref{62} we find energy of
the considered solution as following
\begin{equation}\label{95}
  E= Q(-\partial _{t}) = \frac{1}{4 \pi } \int_{0}^{2 \pi} d \phi \left[ \sigma (\mathcal{L}^{-} + \mathcal{L}^{+}) + \frac{1}{\mu l} (\mathcal{L}^{-} - \mathcal{L}^{+}) \right].
\end{equation}
If we take the limit $\frac{1}{\mu}\rightarrow 0$ this energy reduces to the result presented in \cite{31} exactly. Angular momentum corresponds to the Killing vector $\xi_{(\phi)}=  \partial _{\phi} $. In this case on the background, Eq.\eqref{93} becomes
\begin{equation}\label{96}
  \bar{\lambda} _{\xi _{(\phi)}} - i _{\xi _{(\phi)}} \bar{\omega} = - \bar{e}_{t}.
\end{equation}
Thus, in a similar way, we can find angular momentum of the considered solution as
\begin{equation}\label{97}
  j= Q(\partial _{\phi}) = \frac{l}{4 \pi } \int_{0}^{2 \pi} d \phi \left[ \sigma (\mathcal{L}^{-} - \mathcal{L}^{+}) + \frac{1}{\mu l} (\mathcal{L}^{-} + \mathcal{L}^{+}) \right].
\end{equation}
We can rewrite the metric \eqref{91} as the following form
\begin{equation}\label{98}
  ds^{2}=g_{tt}dt^{2}+g_{\rho \rho} d \rho ^{2} +g_{\phi \phi} d \phi ^{2} +2 g_{t \phi} dt d \phi.
\end{equation}
We have $g_{t \phi} ^{2}- g_{tt} g_{\phi \phi}=0$ on the Killing horizon $H$ which leads to the following equation
\begin{equation}\label{99}
  \mathcal{L}^{+} \mathcal{L}^{-} e^{-2 \rho _{H}} + 4 \mathcal{W}^{+} \mathcal{W}^{-} e^{-4 \rho _{H}} + e^{2 \rho _{H}}= 2 \sqrt{\mathcal{L}^{+} \mathcal{L}^{-}}.
\end{equation}
where we suppose that the Killing horizon is located at $\rho = \rho _{H}$.
In other words, considered solution \eqref{91} may be describes a black hole if $ \rho _{H}$ is a real positive-definite root of the Eq.\eqref{99}. \footnote{As has been discussed in \cite{2} a black hole solution of higher-spin gravity should have 3 properties: 1.have a smooth BTZ limit, 2.have a Lorentzian horizon and a regular Euclidean continuation, 3.have a thermodynamical interpretation. The BTZ black hole is a quotient of $AdS_3$, so it is a solution of higher-spin gravity. Simply one can see that by substituting $\mathcal{W}^{\pm}=0$ into metric \eqref{91}, we obtain BTZ metric. Also in the above we have shown that under some conditions on the functions $\mathcal{L}^{\pm}$ and $\mathcal{W}^{\pm}$, we can find at least one positive real root for $\rho _{H}$. However we are not sure that metric \eqref{91} for non-zero $\mathcal{W}^{\pm}$ can describe a black hole solution of higher-spin gravity. At least we know that these solutions are not black hole with higher spin charges. In order to have a black hole with spin-3 charge, we should extend the $a^{\pm}(x^{\pm})$ in Eq.\eqref{88} to contain the chemical potential conjugate to the $W$ charges. This extension has been done in \cite{13}.}
To show that for some $\mathcal{L}^{\pm}$ and $\mathcal{W}^{\pm}$ such solutions exist, we rewrite Eq.\eqref{99} as
\begin{equation}\label{110}
  e^{3 \rho _{H}} - \sqrt{\mathcal{L}^{+} \mathcal{L}^{-}} e^{ \rho _{H}} \pm 2 \sqrt{- \mathcal{W}^{+} \mathcal{W}^{-}}=0.
\end{equation}
Now, we pick up the plus sign (one can pick up the minus sign and takes the following procedure to find new conditions). One can use the Tschirnhous-Vieta method\footnote{ By using this method, one can find roots of a cubic equation of the form $x^{3}+px+q=0$ as
$$ x_{1} = A \cos \left( \frac{\varphi}{3}\right), \hspace{0.7 cm} x_{2} = A \cos \left( \frac{\varphi + 2 \pi}{3}\right),\hspace{0.7 cm} x_{3} = A \cos \left( \frac{\varphi + 4 \pi}{3}\right),  $$ where $$ A= 2 \sqrt{- \frac{p}{3}}, \hspace{0.7 cm} \varphi = \arccos \left( \frac{3q}{Ap} \right). $$ It is clear that the considered cubic equation has three real roots if $p > 0$.} of solving the cubic equation. Because $ \sqrt{\mathcal{L}^{+} \mathcal{L}^{-}} > 0 $ then Eq.\eqref{110} has three real roots for $e^{ \rho _{H}}$. Also, the following inequality must be satisfied to have at least one real root for $\rho _{H}$
\begin{equation}\label{111}
  0 < \frac{27 \left( - \mathcal{W}^{+} \mathcal{W}^{-} \right)}{ \left( \mathcal{L}^{+} \mathcal{L}^{-} \right)^{\frac{3}{2}} } \leq 1.
\end{equation}
Now it is enough that root to be positive definite. For instance, in the case where $27\left( - \mathcal{W}^{+} \mathcal{W}^{-} \right) \left( \mathcal{L}^{+} \mathcal{L}^{-} \right)^{-\frac{3}{2}} = 1$ we have
\begin{equation}\label{112}
  \rho _{H} = \ln \left( \frac{\left( \mathcal{L}^{+} \mathcal{L}^{-} \right)^{\frac{1}{4}}}{\sqrt{3}} \right) > 0.
\end{equation}
In the above analysis we just consider some special cases. One can extend the range of solutions with a detailed analysis. Thus, for some $\mathcal{L}^{\pm}$ and $\mathcal{W}^{\pm}$, the equation \eqref{99} has real positive definite solutions and we can assume that the solutions \eqref{89} can also describe black holes. The angular velocity of Killing horizon is
\begin{equation}\label{100}
  \Omega _{H} = - \left( \frac{g_{t \phi}}{g_{\phi \phi}} \right) _{H} = \frac{ ( \sqrt{\mathcal{L}^{-}} - \sqrt{\mathcal{L}^{+}} ) }{ l (\sqrt{\mathcal{L}^{-}} + \sqrt{\mathcal{L}^{+}})}.
\end{equation}

 The angular velocity of Killing horizon is
\begin{equation}\label{100}
  \Omega _{H} = - \left( \frac{g_{t \phi}}{g_{\phi \phi}} \right) _{H} = \frac{ ( \sqrt{\mathcal{L}^{-}} - \sqrt{\mathcal{L}^{+}} ) }{ l (\sqrt{\mathcal{L}^{-}} + \sqrt{\mathcal{L}^{+}})}.
\end{equation}
Now, we want to find the entropy of the considered black hole solution \eqref{91} by the entropy formula \eqref{86}. By substituting $\omega _{\phi}$ from Eq.\eqref{90} into the formula \eqref{86} we obtain
\begin{equation}\label{101}
  S = 2 \int_{0}^{2 \pi} \frac{1}{\sqrt{ g_{\phi \phi}}}  \left( \sigma  g _{\phi \phi} - \frac{1}{\mu} g _{ t \phi} \right) d \phi,
\end{equation}

where have we used Eq.\eqref{9}. By considering the explicit form of the metric and by using Eq.\eqref{99}, the above equation becomes
\begin{equation}\label{102}
  S = 2 l \int_{0}^{2 \pi} d \phi \left[ \sigma (\sqrt{\mathcal{L}^{-}} + \sqrt{\mathcal{L}^{+}})  + \frac{1}{\mu l} (\sqrt{\mathcal{L}^{-}} - \sqrt{\mathcal{L}^{+}}) \right].
\end{equation}
Since the BTZ black hole solution can be described by the connections of the form Eq.\eqref{87}, we can find simply the energy,
angular momentum, and entropy of this black hole by our obtained formula. But we can not use these connections for
higher-spin black hole solutions introduced in \cite{13, 34',2}. One can see the expression $a^{\pm}(x^{\pm})$ for
these class of higher-spin black holes in \cite{31,32}, where the authors have found the energy and entropy of them.
By inserting following functions into metric Eq.\eqref{91}, we obtain a metric for BTZ black hole of mass $M$ and angular momentum $a$
\begin{equation}\label{103}
  \mathcal{L}^{-} = \frac{2}{l} (Ml+a), \hspace{7 mm} \mathcal{L}^{+} = \frac{2}{l} (Ml-a),\hspace{7 mm} \mathcal{W}^{\pm}=0,\hspace{7 mm} .
\end{equation}
In this case, equations \eqref{95}, \eqref{97} and \eqref{102} reduce to following quantities, respectively
\begin{equation}\label{104}
  E = M + \frac{a}{\mu l^2},
\end{equation}
\begin{equation}\label{105}
  j = a + \frac{M}{\mu},
\end{equation}
and
\begin{equation}\label{106}
  S = 4 \pi \left( r_{+} + \frac{1}{\mu l} r_{-} \right).
\end{equation}
These results are what we expected in the BTZ case. In the equation \eqref{106}, $r_{-}$ and $r_{+}$ are respectively inner and outer horizon radii of BTZ black hole,
\begin{equation}\label{107}
  r_{\pm} = \sqrt{2 \left( M \pm \sqrt{M^2 l^2 - a^2} \right)} .
\end{equation}
The horizon angular velocity Eq.\eqref{100} in this case becomes $\Omega _{H} = \frac{r_{-}}{l r_{+}} $, also the surface gravity is
\begin{equation}\label{108}
  \kappa = \left[ \sqrt{-\frac{1}{2} \nabla ^{\mu} \xi ^{\nu} \nabla _{\mu} \xi _{\nu} } \right]_{H} = \frac{ r_{+}^{2} - r_{-}^{2} }{l^2 r_{+}},
\end{equation}
where $\xi$ is the Killing horizon generator vector field. It is worth to say that equations \eqref{104}-\eqref{108} satisfy the first law of black hole mechanics, that is
\begin{equation}\label{109}
  \delta E = T_{H} \delta S + \Omega _{H} \delta j,
\end{equation}
where $T_{H} = \kappa / (2 \pi) $.

\section{Conclusion}
The spin-3 gravity is covariant under generalized Lorentz gauge transformation as well as diffeomorphism. Using this fact, one can define quantities (such as generalized torsion 2-form Eq.\eqref{15} and generalized curvature 2-form Eq.\eqref{16}) so that they are covariant under generalized Lorentz gauge transformation. We saw that inorder to define an exterior covariant derivative, we need to introduce a spin-connection which transform under
generalized Lorentz gauge transformations as Eq. \eqref{13}. The ordinary Lie derivative is not covariant under this transformation so we defined the generalized Lie derivative Eq.\eqref{19} so that it is covariant under generalized Lorentz gauge transformations. We showed that by demanding torsion-free condition and choosing $\lambda _{\xi} = i_{\xi} \omega$, this formalism reduces to ordinary one which is describes by the Chern-Simons action \eqref{28}  with gauge group $SL(3,\mathbb{R}) \times SL(3,\mathbb{R})$. In the section 3, we extended our study from spin-3 gravity theory to the topologically massive spin-3 gravity. Presence of the Chern-Simons topological term makes the Lagrangian of this theory, Eq.\eqref{34}, non-invariant under generalized Lorentz gauge transformations. To obtain the conserved charges of the topologically massive spin-3 gravity therefore we should use the method of obtaining conserved charges in non-covariant theories which firstly was introduced by Tachikawa for ordinary topological massive gravity \cite{33}. Also, we showed that all the solutions of the spin-3 gravity are solutions of the spin-3 topologically massive gravity. In section 4, we found a general formula to calculate conserved charge of the spin-3 topologically massive gravity corresponds to a Killing vector field $\xi$. We have defined the off-shell ADT current \eqref{59} associated to a vector field $\xi$ by virtue of the Bianchi identities \eqref{55}, then we showed that this current is conserved when $\xi$ is a Killing vector field. Hence, using Poincare lemma, we defined off-shell ADT conserved charge \eqref{61} associated to a Killing vector field $\xi$. Consequently, by integrating off-shell ADT conserved charge, we obtained quasi-local conserved charge \eqref{62}. It should be noted that this result is independent of the integration surface $\Sigma$ and then we can use the formula \eqref{62} to calculate the conserved charge of solutions which are not asymptotically AdS. We showed that our general formula Eq.\eqref{72} reduces to the previous one for the spin-3 gravity presented in \cite{31} when we take into account only transformation under diffeomorphism, without considering generalized Lorentz gauge transformation (i.e. $\lambda _{\xi} =0$), and by taking $\frac{1}{\mu}\rightarrow 0$. In the section 5, we fixed generator of Lorentz gauge transformation $\lambda _{\xi}$ by demanding that generalized Lie derivative of the generalized dreibein $e_{\mu}$ becomes zero when $\xi$ is Killing vector field and $\lambda _{\xi}$ transforms as \eqref{20} under generalized Lorentz gauge transformation. In the section 6, we obtained a general formula \eqref{86} for the entropy of black hole solutions of the spin-3 topologically massive gravity. In the section 7, we applied this formalism to calculate energy, angular momentum and entropy of a special black hole solution and we have shown that obtained results are consistent with previous results in the limiting cases. Moreover our result for energy, angular momentum and entropy are consistent with the first law of black hole mechanics.
\section{Acknowledgments}
M. R. Setare thank Mu-In. Park and Bin. Chen for helpful comments and discussions. We also thank the referee for his/her
useful comments, which assisted in preparing a better frame for this study.
\appendix

\section{$ SL(3,\mathbb{R})$ generators}
In the section 7, we used the following basis of $ SL(3,\mathbb{R})$ generators \cite{1,13}
\begin{equation}\label{A1}
  \begin{split}
       & L_{1}=\left(
                 \begin{array}{ccc}
                   0 & 0 & 0 \\
                   1 & 0 & 0 \\
                   0 & 1 & 0 \\
                 \end{array}
               \right)
        \hspace{0.5 cm}
        L_{0}=\left(
          \begin{array}{ccc}
            1 & 0 & 0 \\
            0 & 0 & 0 \\
            0 & 0 & -1 \\
          \end{array}
        \right)
        \hspace{0.5 cm}
        L_{-1}=\left(
          \begin{array}{ccc}
            0 & -2 & 0 \\
            0 & 0 & -2 \\
            0 & 0 & 0 \\
          \end{array}
        \right)
        \\
       &W_{2}=\left(
          \begin{array}{ccc}
            0 & 0 & 0 \\
            0 & 0 & 0 \\
            2 & 0 & 0 \\
          \end{array}
        \right)
         \hspace{0.5 cm}
         W_{1}=\left(
          \begin{array}{ccc}
            0 & 0 & 0 \\
            1 & 0 & 0 \\
            0 & -1 & 0 \\
          \end{array}
        \right)
        \hspace{0.5 cm}
         W_{0}=\frac{2}{3}\left(
          \begin{array}{ccc}
            1 & 0 & 0 \\
            0 & -2 & 0 \\
            0 & 0 & 1 \\
          \end{array}
        \right)
        \\
       & W_{-1}=\left(
          \begin{array}{ccc}
            0 & -2 & 0 \\
            0 & 0 & 2 \\
            0 & 0 & 0 \\
          \end{array}
        \right)
        \hspace{0.5 cm}
         W_{-2}=\left(
          \begin{array}{ccc}
            0 & 0 & 8 \\
            0 & 0 & 0 \\
            0 & 0 & 0 \\
          \end{array}
        \right).
  \end{split}
\end{equation}
The generators satisfy the following commutation relations
\begin{equation}\label{A2}
  \begin{split}
       & [L_{i},L_{j}]=(i-j) L_{i+j} \\
       & [L_{i},W_{m}]=(2i-m) W_{i+m} \\
       & [W_{m},W_{n}]=-\frac{1}{3}(m-n)(2m^{2}+2n^{2}-mn-8)L_{m+n}.
  \end{split}
\end{equation}
where $-1 \leq i,j \leq 1 $ and $-2 \leq m,n \leq 2$. Also, the non-zero traces are
\begin{equation}\label{A3}
  \begin{split}
       & tr(L_{0}L_{0})=2, \hspace{0.5 cm} tr(L_{1}L_{-1})=-4 \\
       & tr(W_{0}W_{0})=\frac{8}{3}, \hspace{0.5 cm} tr(W_{1}W_{-1})=-4,\hspace{0.5 cm} tr(W_{2}W_{-2})=16 .
  \end{split}
\end{equation}

\end{document}